\documentclass[12pt]{article}
\usepackage{graphicx}
\usepackage{amssymb}
\usepackage{amsmath}
\usepackage{xcolor} 
\usepackage{float} 

\usepackage{cancel} 
\usepackage{ulem} 

\usepackage{braket}

\usepackage{bm}

\newcommand{\beq}{\begin{equation}}

\newcommand{\eeq}{\end{equation}}
\newcommand{\bea}{\begin{eqnarray}}
\newcommand{\eea}{\end{eqnarray}}

\newcommand{\oh}{\frac{1}{2}}

\newcommand{\HE}{{\rm H\! -\!E}}

\begin{document}

\begin{center}
\vspace{24pt}
{ \large \bf Pseudo-Cartesian coordinates in a model of Causal Dynamical Triangulations}

\vspace{30pt}

{\sl J. Ambj\o rn}$\,^{a,b}$,
{\sl Z. Drogosz}$\,^{c}$,
{\sl J. Gizbert-Studnicki}$\,^{c}$,
{\sl A. G\"{o}rlich}$\,^{c}$,
{\sl J. Jurkiewicz}$\,^{c}$

\vspace{24pt}
{\footnotesize

$^a$~The Niels Bohr Institute, Copenhagen University\\
Blegdamsvej 17, DK-2100 Copenhagen \O , Denmark.\\
{ email: ambjorn@nbi.dk}\\

$^b$~IMAPP, Radboud University, \\
Nijmegen, PO Box 9010, The Netherlands

\vspace{10pt}

$^c$~The M. Smoluchowski Institute of Physics, Jagiellonian University,\\
Lojasiewicza 11, PL 30-348, Krakow, Poland.\\
{ email: zbigniew.drogosz@uj.edu.pl, jakub.gizbert-studnicki@uj.edu.pl, andrzej.goerlich@uj.edu.pl, jerzy.jurkiewicz@uj.edu.pl}\\

\vspace{10pt}

}
\vspace{48pt}

\end{center}


\begin{center}
{\bf Abstract}
\end{center}

Causal Dynamical Triangulations is a non-perturbative quantum gravity model, defined with a lattice 
cut-off. The model can be viewed as defined with a proper time but with no reference 
to any three-dimensional spatial background geometry. It has  four phases, depending on the parameters 
(the coupling constants) of the model. The particularly interesting behavior is observed in the so-called de Sitter phase, where the spatial three-volume 
distribution as a function of proper time has a semi-classical
behavior which can be obtained from an effective mini-superspace action. 
In the case of the three-sphere spatial topology, it has been difficult to extend the 
effective semi-classical description in terms of proper time and
spatial three-volume to include genuine spatial coordinates, partially because of the 
background independence inherent in the model. However,
if the spatial topology is that of a three-torus, it is possible to define a number of new 
observables that might serve as spatial coordinates as well as new observables related to
the winding numbers of the three-dimensional torus.
The present paper outlines how to define the observables,  
and how they can be used in numerical simulations of the model.
\newline


\newpage

\section{Introduction}

The model of Causal Dynamical Triangulations (CDT) in four dimensions was originally formulated for systems where
the spatial topology of the Universe was assumed to be spherical ($S^3$). In most cases the (Euclidean) proper time
was assumed to be periodic with some period $T$. The spherical topology of the spatial part of the Universe 
gives a number of advantages, among which the main one is its relative simplicity. 
Unfortunately it also has disadvantages. In the original model the path integral includes a summation  
over all spatial simplicial geometries corresponding to  abstract triangulations of $S^3$.
This background independence makes it almost impossible to define reference points, except in the time direction.
Another problem is caused by the numerical algorithm used in the computer simulations, 
which made the most interesting range in the coupling constants, i.e.,
the one where the phase transition lines meet, practically inaccessible. 
These two deficiencies may be overcome using a different
choice of the spatial topology, namely the toroidal topology ($T^3$). In this case, the path integral defining the model 
will include a summation over abstract triangulations of $T^3$.  

CDT with a toroidal spatial topology was shown to have a phase structure similar to that observed when the 
spatial topology was spherical  \cite{torus1}. We have a system with four distinct phases,
where the phase transition lines are approximately the same as in 
the spherical case\footnote{Some differences may be attributed to relatively large finite size effects.}. 
The most interesting range in the coupling constants space is now, surprisingly, fully 
accessible\footnote{The detailed analysis of this range will be the subject of an independent publication.} and shows that the phase structure is more complicated than that conjectured earlier.
Instead of a quadruple point\footnote{It should be emphasized that the existence of a quadruple point was only 
a conjecture in the spherical case since it was located in the region of coupling constant space that was inaccessible to computer simulations.} we seem to observe two triple points, connected by a phase transition line.   

A finite system with a toroidal $T^3$ geometry can be viewed as an infinite system, where a finite elementary cell is
periodically repeated infinitely many times in three directions. The resulting copies of the elementary cell may be numbered by the winding numbers. The boundary of the elementary cell is not uniquely defined. 
However, as we will show below, it is possible to introduce additional constraints that can 
make such a definition unique.

In this article we will discuss the possibility of using the set of boundaries as a reference frame that 
allows us to define a set of pseudo-Cartesian coordinates in the elementary cell. 
The spatial volume distribution associated with these 
coordinates can be interpreted as a distribution of $\sqrt{g(x,y,z,t)}$, which potentially
permits measuring the effective action parametrized not only by a time variable but also by spatial variables.
 We will also define a set of new observables
related to the topologically closed geodesic loops with non-trivial winding numbers.

\section{Toroidal topology}  

Details of a formulation of the CDT model were discussed earlier in a number of articles (for a review see e.g. \cite{report}). Let us 
mention here a few essential points. The basic idea is to consider the analogue of the Feynman path integral
for the amplitude between two spatial states with initial and final three-geometries. 
 It is assumed that 
trajectories satisfy a condition of {\it causality}, which means that the considered spacetimes 
are globally hyperbolic 
(permitting a global time foliation) and that the spatial topology is fixed on all leaves of the foliation.
We use the Regge-type discretization of the spacetime \cite{regge}. The spacetime is constructed from elementary
building blocks -- four-dimensional simplices -- with a unique length of spatial edges and a unique length
of time edges. Each vertex in the system has a uniquely defined
integer time parameter. As a consequence, we have two types of four-simplices: the $\{4,1\}$-simplices (with four vertices at
time $t$ and one at $t\pm 1$) and $\{3,2\}$-simplices (with three vertices at time $t$ and two at $t\pm 1$). The simplices are 
glued together along three-dimensional {\it tetrahedral faces} to form a {\it simplicial manifold} 
with additional regularity constraints imposed (all simplices and sub-simplices with a particular set of vertex labels appear at most 
once). The existence of a global foliation means that each configuration can be analytically continued between the
Lorentzian and Euclidean geometry. The Wick rotation can be interpreted as analytic continuation of the 
length of time links. In the Euclidean formulation the memory of the time orientation is preserved. In most cases studied
the systems are assumed to be periodic in the (Euclidean) time, which means that we do not need to specify initial and
final spatial geometric configurations.

For every configuration of the system we may define a dual lattice, where each simplex becomes a dual vertex
and each face becomes  a dual link, etc. On a dual lattice, from each vertex emerge exactly five links, corresponding to exactly five neighbors of a simplex on a direct lattice. For simplicity, we will assume that all  links on a dual lattice have the same length. The basic observable to be
used will be the length of a geodesic line between two simplices, i.e., the minimal number of steps (links on the dual lattice) necessary to connect
the two simplices. The geodesic line is usually not unique, but the minimal distance between simplices is
always well defined. 

A spacetime trajectory $\cal{T}$ appearing in the path integral   is weighted by  the 
exponential of (minus) the Hilbert-Einstein action, which in Euclidean formulation
becomes real and can be interpreted as a probability
\begin{equation}
{\cal{P}}({\cal T}) \propto \  e^{-S_{\HE}({\cal T})}. 
\end{equation}
For a system with a finite number of simplices the action $S_{\HE}({\cal T})$ takes a particularly simple form\footnote{
For a piecewise linear four-dimensional geometry, defined by a four-dimensional triangulation, 
Regge \cite{regge}  provided a simple geometric expression for the Hilbert-Einstein action, expressed
in terms of link lengths and deficit angles of triangles in the triangulation. In our case, this expression 
simplifies further since our link lengths take only two values, corresponding to spacelike and timelike links. 
The end result is that the action can be expressed entirely in terms of the number of vertices $N_0$,  the number 
of four-simplices  $N^{\{4,1\}}$ and $N^{\{3,2\}}$, and the ratio between the length of spacelike and timelike links.
See \cite{report} for details.}
\begin{equation}
S_{\HE}({\cal T}) = -(K_0+6\Delta) N_0 + K_4 \left( N^{\{4,1\}}+ N^{\{3,2\}}\right)+\Delta  N^{\{4,1\}}.
\label{Sdisc}
\end{equation}
In (\ref{Sdisc}) the global numbers $N_0$, $N^{\{4,1\}}$ and $N^{\{3,2\}}$ denote correspondingly the number of vertices
and the numbers of simplices of each of the two types in a triangulation ${\cal T}$. The dimensionless coupling constants
are $K_0$ -- related to the inverse of the gravitational constant, the cosmological constant $K_4$ and $\Delta$ -- 
a function of the ratio between the time and spatial length of edges. Typically in a configuration $N_0 \propto
N_4= N^{\{4,1\}}+ N^{\{3,2\}}$. Different triangulations correspond in general to different piecewise linear 
geometries, and the effective number of triangulations parametrized by the same set of global numbers 
$N^{\{4,1\}}$ and $N^{\{3,2\}}$ at $K_0,~\Delta$
(i.e., with the same action) grows in the leading order as
\begin{equation}
{\cal{N}}(N_4;K_0,\Delta)\approx e^{K_4^{crit}(K_0,\Delta) N_4}.
\end{equation}
 As a consequence, the amplitude is defined only
for $K_4 > K_4^{crit}(K_0,\Delta)$ and the limit $K_4 \to K_4^{crit}(K_0,\Delta)$ is the large-volume limit where the average number
of simplices becomes large and where we may try to define a continuum limit.
\
One can define similar models in 1+1 dimensions and 2+1 dimensions. Only in 1+1 dimensions can
the model be solved analytically \cite{two-d}. In 2+1 dimensions there exist partial analytic results \cite{three-d}, but to achieve 
 a reliable understanding
 of the phase structure one has to rely on Monte Carlo simulations.   
In 3+1 dimensions  Monte Carlo simulations are the only tool available \cite{four-d}.
In all the simulations of a 3+1 system we use a set of 7 local {\it moves}, which
preserve topology and foliation and which are ergodic in the class of triangulations we use (see \cite{report} for 
details).  The Monte Carlo evolution is performed at a given set of bare couplings $K_0$ and $\Delta$. The limit
$K_4 \to K_4^{crit}$ is studied by considering a growing sequence of systems with  
volumes $\bar{N}^{\{4,1\}}$, where the volume
$N^{\{4,1\}}$ is forced to fluctuate around $\bar{N}^{\{4,1\}}$. The initial configuration with a prescribed spatial 
topology $\Sigma$ and the time extent $T$ fixes the topology of the studied system to be $\Sigma\times  T^1$. 
The computer program
works for arbitrary spatial topology $\Sigma$ and arbitrary $T$, so the correct choice of the initial configuration with the desired spatial topology $\Sigma$ is 
a very important element of each simulation, although different choices with the same topology  give
equivalent results.

The choice of an initial  toroidal spatial geometry $\Sigma =  T^3$ 
was discussed in our earlier paper \cite{torus2}. The initial configuration used to start a Monte Carlo simulation is not a 
minimal configuration (in the sense discussed in \cite{torus2}), but it is a very convenient choice 
from the point of view of  the present article. 
The configuration is based on
a triangulation of a four-dimensional hypercube \cite{cube}. The topological hypercube is divided into 16 four-simplices,
out of which 10 are of the $\{4,1\}$ type and 6 of the $\{3,2\}$ type. To build a configuration with a $T^3\times T^1$ topology, one needs a number of hypercubes
that is even and greater or equal to four. The last
restriction comes from the manifold conditions: the two vertices of a link must be different.
The smallest such configuration, 
assuming that the time period is $T=4$,
has $N_4=16\cdot 4^4=4096$ simplices. If $\Sigma$ has the topology of $S^3$ one can use a much smaller starting 
configuration.

A system with $T^3$ spatial topology can be viewed as an infinite system, where a finite size {\it elementary cell} is
periodically repeated in all spatial directions (for periodic b.c. this is also the case in the time direction). 
The starting configuration described above corresponds to a system composed of periodically 
repeated triangulated four-cubes with $N_4$ simplices each.
The definition of an elementary cell is not unique. 
We may additionally request that the boundary between the neighboring cells
has a minimal three-volume. This condition, for a starting configuration, does not lead to a unique
choice: in fact, in every spatial direction we may choose the {\it initial} position of a boundary in
four equivalent ways, which translates into $4^3$ equivalent spatial boundaries. This is a result of the relatively high symmetry of the initial configuration, which is invariant under
time and spatial shifts by one unit. The exact shape of each boundary may be locally modified in such a way that
it remains simply connected. This, in general, changes the three-volume of the boundary. To control the position of  boundaries,
we introduce in the coding of the geometry an additional information for each interface 
between neighboring simplices (i.e., for each dual link).
In the four-dimensional case each simplex has five neighbors, as explained above. 
For each dual link we introduce the integer-valued spatial vector
$\vec{v}$, with entries having values $\pm 1,~0$ in each spatial direction. Value 0 corresponds to the case where both
the simplex and a particular neighbor are in the same elementary cell. Values $\pm 1$ mean that crossing the interface
we enter the neighboring elementary cell in the positive or the negative direction. For the initial configuration choice
we assign values to the vector field $\vec{v}$ for all interfaces between neighboring simplices.
In any given configuration, we may form a sum of vector fields along any loop joining the simplices on a dual lattice. For a trivial closed loop the sum must be a zero vector. For a non-trivial
closed loop (closed by periodic b.c.) the sum gives topological information about the loop's winding number.

 In the numerical simulations the standard algorithm is using a 
finite set of local {\it moves}. Each move affects only a finite
part of the configuration (i.e., of the triangulation), keeping the rest  unchanged. Since Hilbert-Einstein action (\ref{Sdisc}) does not depend on the position of the boundaries, the decision to accept (or reject)
the move does not depend on a position of the boundaries. The position of the boundaries will
however play an important role in the construction proposed in this paper. Therefore, before the move is performed we
check if  the boundaries between elementary cells
pass through the affected region of the configuration. If this is not the case, the move is performed 
using the standard Monte Carlo algorithm.
If, however, a boundary between two cells crosses the part of the triangulation where 
the geometry of connections between simplices
is to be changed, then we modify locally  the position of the boundary. The modification
is  done in such a way that the boundary remains connected, but is deformed so as to lay wholly
outside of the region of the triangulation that is going to be affected by the move.
Such a modification can always be done and  does not change the action, but  
in general it will increase the size of the boundary. The move can now be performed
in a standard way. 
After the move, we check if a simple local modification of the boundary can reduce its size
(its three-volume).
If this is the case, the modification of the position of the boundary is made.
From time to time,we perform an additional check, independent of the moves, whether a local modification of any part of the boundaries can minimize their volumes. If so, then such a modification is always accepted.

All measurements described in this article were performed for the toroidal spatial topology in the de Sitter phase (the so-called C phase) at a special
point in the parameter space $K_0=2.2$ and $\Delta=0.6$ for systems with $T=4$ and a sequence
of volumes $N^{\{4,1\}}$. The same point in the parameter space was earlier analyzed in simulations of systems 
where the configurations had spherical spatial topology \cite{four-d}, as well as in systems with configurations
of  toroidal spatial topology \cite{torus2}.

\section{Pseudo-Cartesian coordinates}

\begin{figure}[H]
\begin{center}
{\includegraphics[width=11cm]{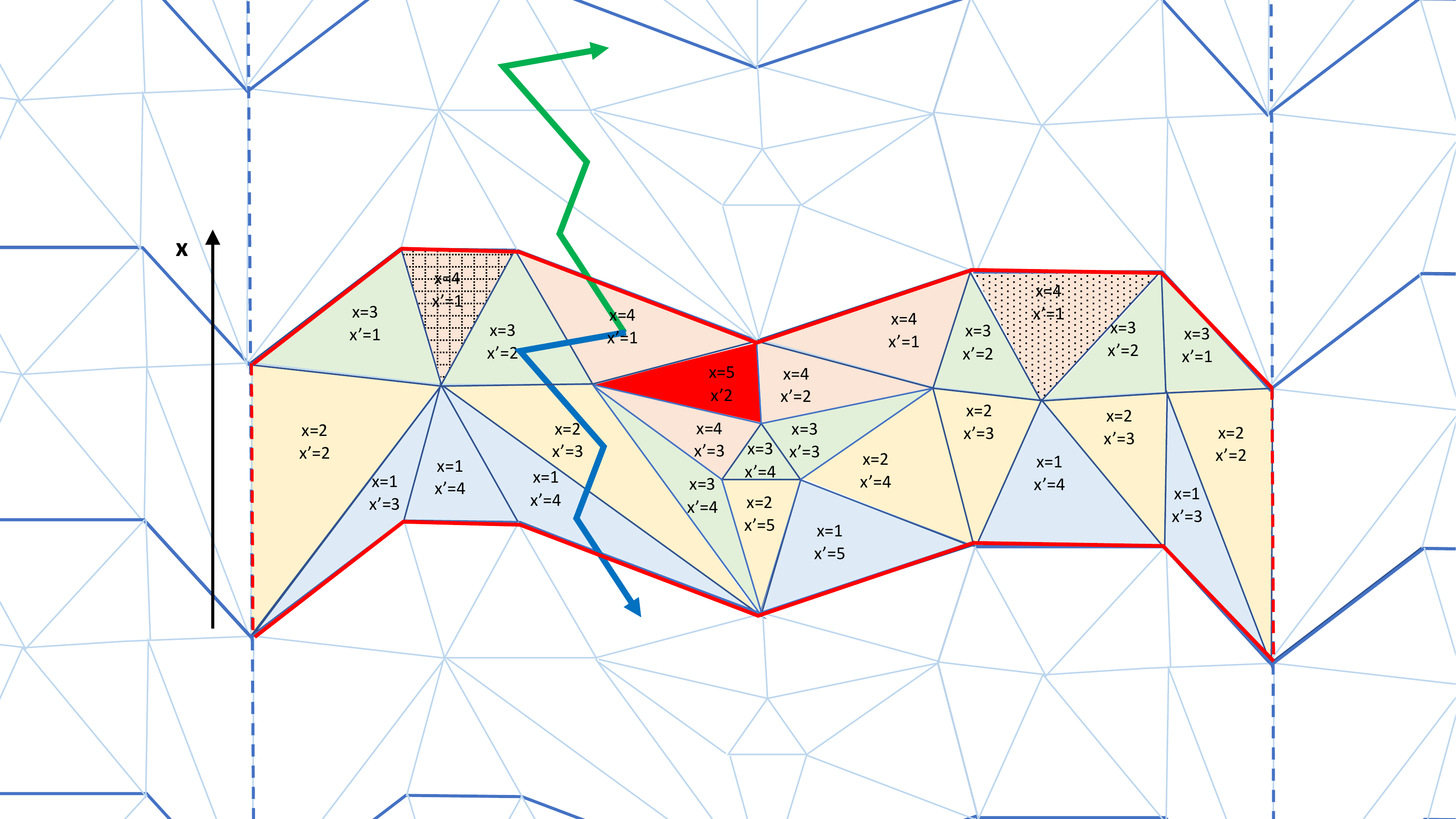}} 
\end{center}
\caption{{The (two dimensional) visualization of a triangulation (colored triangles) with toroidal topology. The (smallest) boundary orthogonal to the   ${\bf x}$ direction is plotted  as a red solid line, and the boundary orthogonal to the other direction as a red dashed line (the red solid lines, as well as the red dashed lines, are identified). 
Different colors mark  different ${\bf x}$ coordinates. All triangles of the same color and texture  form single {\it slices} (see Section \ref{sec:4} for description): for   ${\bf x}= 1, 2, 3, 5$ one has only a single slice for each  ${\bf x}$ coordinate, but for ${\bf x}=4$ one can distinguish three separate slices. All triangles, but the dark red one, belong to the {\it trunk}, and the dark red triangle belongs to the {\it branch} (see Section \ref{sec:4} for description).
Alternatively (see Section \ref{sec:5}), one can treat the toroidal triangulation as an elementary cell which  periodically repeats  in each direction (blank triangles). For each triangle  one can find (one or more)  minimal loops with a nontrivial winding number: all triangles whose centers are marked by a solid zigzag arrow belong to the same $\{1,0\}$  loop (green arrow), and also to the same  $\{-1,0\}$  loop (blue arrow), the length of these loops is $4$. One can construct similar loops for all other triangles and measure their length. One can also construct the  minimal loops with other winding numbers in all directions.} }\label{fig:visualize}
\end{figure}

A given set of boundaries can be used to define what we will denote as the {\it pseudo-Cartesian coordinates}. 
The procedure is as follows:
\begin{itemize}
\item We start from a boundary\footnote{The boundaries are labeled 
$x,y$ and $z$, based on their positions in the initial
configuration, which is related to the hypercube.}, say, orthogonal to the direction ${\bf x}$, and we mark all simplices adjacent to the boundary
in the direction we define to be {\it positive} as having the coordinate ${\bf x} = 1$. 
\item We move (on the dual lattice) in the positive direction to the second layer of simplices at a unit distance from the first layer.
We mark these simplices as having a coordinate ${\bf x}=2$.
\item We continue the same process until all simplices in the system are marked, and a maximal extent in the $x$ direction is
reached.
\item The same procedure can be started from the layer of simplices adjacent to a boundary in the {\it negative}
direction. The simplices in that layer are marked as having a coordinate ${\bf x'}=1$.
\item We continue the process as before.
\item This way each simplex is assigned the values of the coordinates ${\bf x}$ and ${\bf x'}$.
\item Exactly the same method can be used to define the coordinates ${\bf y}$, ${\bf y'}$, ${\bf z}$ and ${\bf z'}$.
\end{itemize} 
{For a (two-dimensional) visualization of the procedure please refer to Figure \ref{fig:visualize}.}
\newline

One should note that although each three-dimensional boundary is simply connected, this does not need to be the case
for three-dimensional surfaces separating ${\bf x}$ and ${\bf x} +1$ layers. The set of numbers $\{{\bf x},~{\bf x'}\}$ 
represents the distance of a simplex to a boundary, i.e., the number of links on the dual lattice, in negative and positive directions.

\begin{figure}[H]
\begin{center}
{\includegraphics[width=12cm]{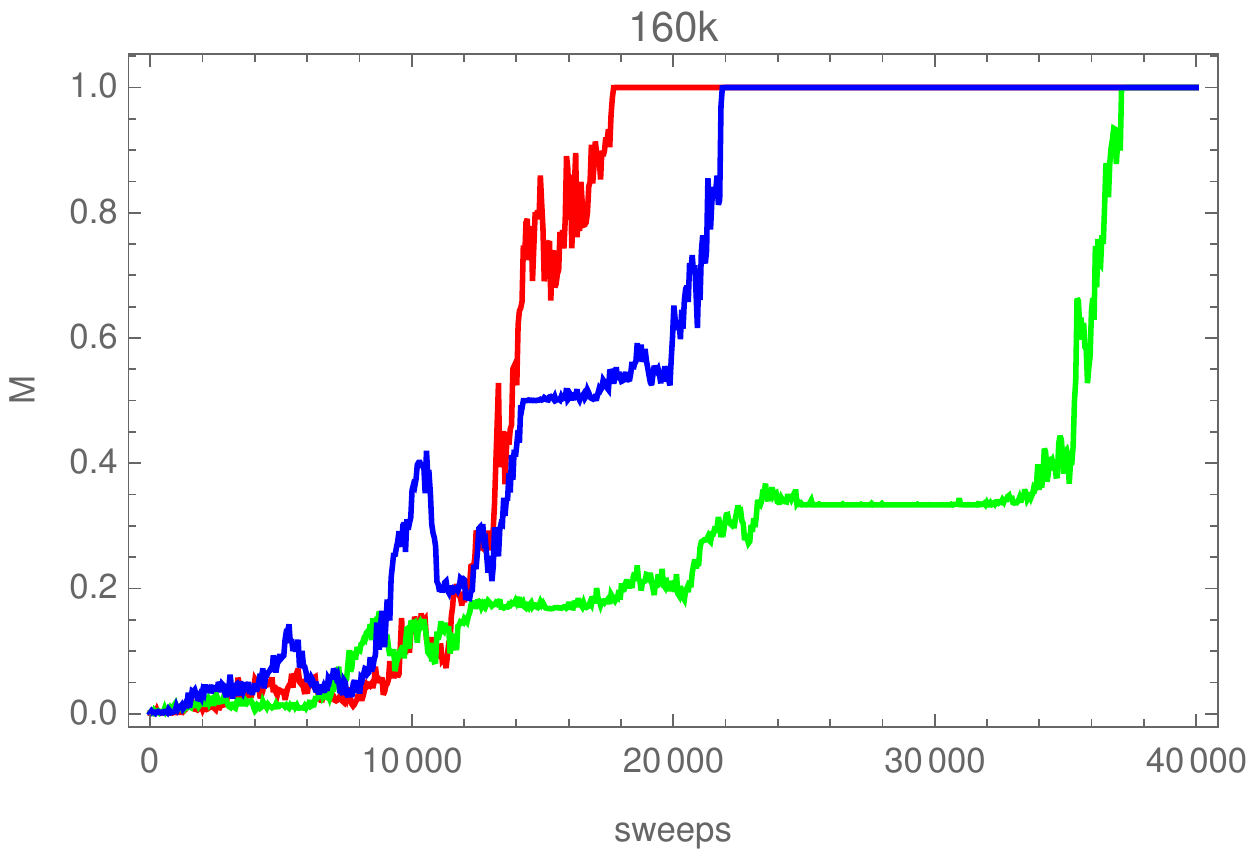}} 
\end{center}
\caption{The example of the evolution in computer time of the  boundary merging ratio $M$ in
 all three directions (${\bf x}$ â red line, ${\bf y}$ â green line, ${\bf z}$ â blue line) for a system with a volume $N^{\{4,1\}}=$160k. The two initial choices of the
 position of boundaries are the most distant in each direction.}\label{fig:merg}
\end{figure}

The first question one may ask is if the definition of a boundary between elementary cells, 
as implemented  by the minimization procedure, is unique. 
For the simple initial configuration described above this is definitely not the case. As already explained, in each spatial
direction we may choose the initial position of a boundary between the copies of the elementary cell in four 
equivalent ways, giving rise to $4^3$ equivalent (minimal) boundaries. Since the updates of geometry do not ``feel'' the position of the boundaries, 
we can run a parallel simulation,
where the evolution of geometry is exactly the same, but the initial choice of a position of a boundary is different.
In this way, during the thermalization process, each simplex can be assigned its coordinates using the two different
positions of a boundary. Comparing the two coordinates, say ${\bf x}$ and ${\bf \tilde{x}}$, we define the difference
between the two alternative definitions by measuring the ratio $M$ of the number of simplices for which 
${\bf x}={\bf \tilde{x}}$ to the total number of simplices. This quantity is initially zero. A value $M=1$ means
that two definitions become equivalent, or in other words, that the two boundaries fully overlap. Analogous quantities can
be defined in ${\bf y}$ and ${\bf z}$ directions. In the figure \ref{fig:merg}
we present the evolution of
the parameter $M$ for three directions in a system with $N^{\{4,1\}}=$160k. The initial position of boundaries
was chosen as the most distant permitted by the symmetry of the starting configuration. Similar experiments
were repeated for other initial choices of boundaries and other system sizes. In all cases the result was the same,
indicating that the algorithm used is powerful enough to find a global minimum of the boundary size in all
directions. It is a non-trivial result, because na\"ively one
may expect the existence of a complicated landscape of local minima. This also means that we do not have to 
worry about the initial choice of the position of boundaries.

Since the considered systems are periodic in (Euclidean) time with a period $T$, each simplex can be assigned a
coordinate ${\bf t}$. The layer with ${\bf t}=1$ is chosen as the set of $\{4,1\}$-simplices with four vertices
at $t=1$. We follow the same prescription as before, for instance ${\bf t}=2$ is composed of a set of $\{3,2\}$-simplices
at a unit distance from the ${\bf t}=1$ in the {\it positive} time direction. We continue until we reach the layer ${\bf t}=4T$ formed
from $\{1,4\}$ simplices. We could define coordinates ${\bf t'}$ similarly as before, but for the time
direction for each system we have a trivial relation ${\bf t}+{\bf t'}=4T+1$, resulting from the trivial periodicity
by construction in the time direction.

\begin{figure}[H]
\begin{center}
{\includegraphics[width=12cm]{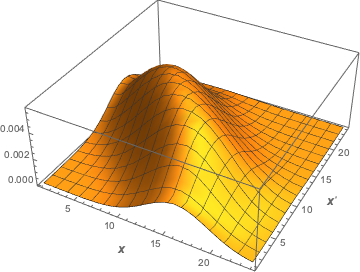}} 
\end{center}
\caption{The  distribution  $P({\bf x},{\bf x'})$ for a system with $N^{\{4,1\}}=$160k.   }
\label{xxprim}
\end{figure}

In the spatial directions a similar relation is not satisfied 
and the distribution of values for, say, ${\bf x}$ and ${\bf x'}$ is  non-trivial. 
Below, in the figure \ref{xxprim}, we present the probability distribution $P({\bf x},{\bf x'})$
to find a simplex with given values of coordinates  ${\bf x}$ and ${\bf x'}$ in a system with $N^{\{4,1\}}=$160k.
The distribution is constructed by averaging  over  800 statistically independent configurations and summed over all 
simplices in the system.

\begin{figure}[H]
\begin{center}
{\includegraphics[width=7cm]{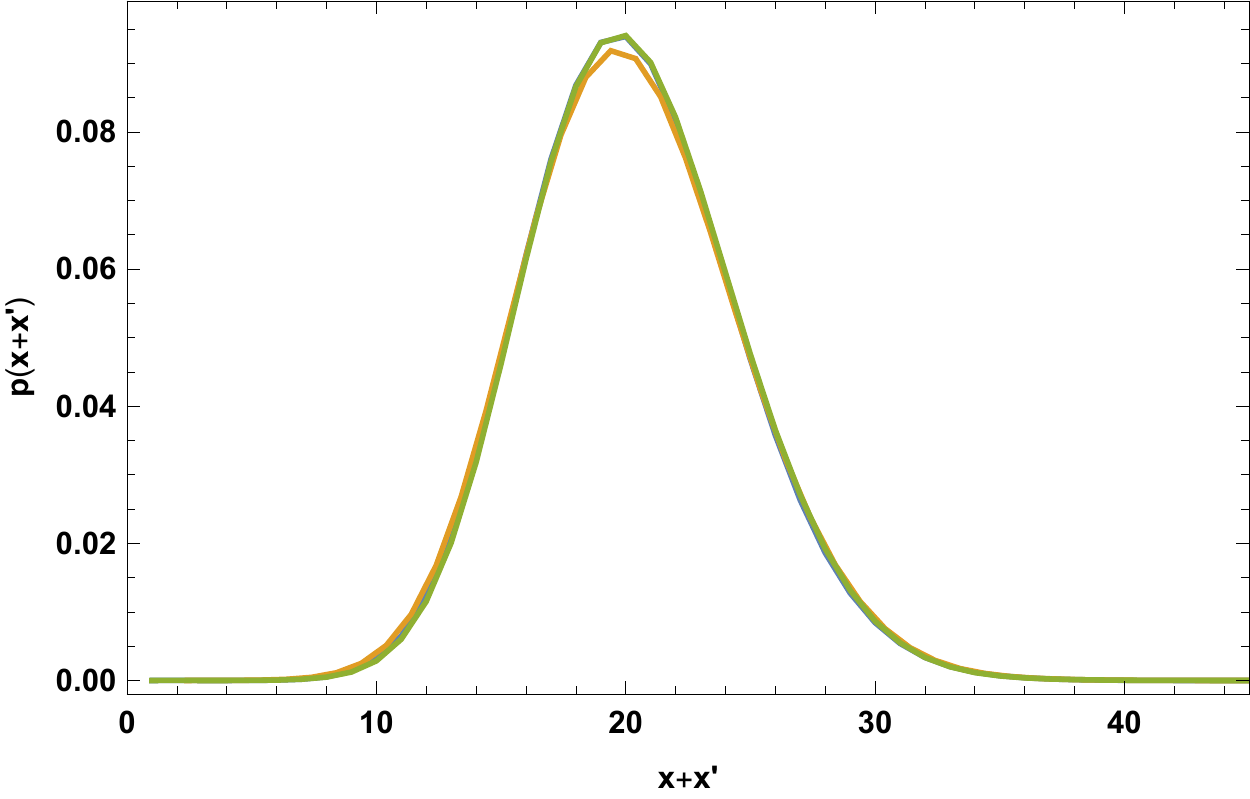}} 
{\includegraphics[width=7cm]{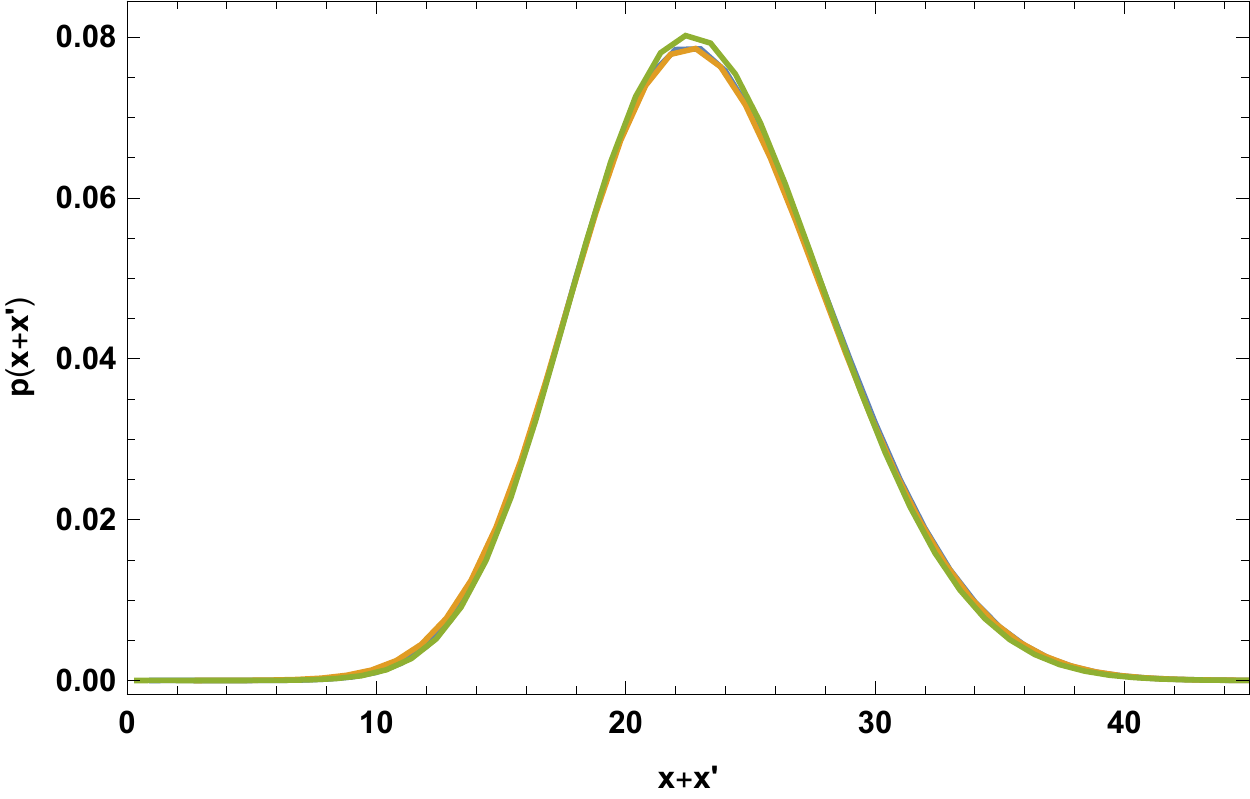}} 
\end{center}
\caption{Distributions of $p({\bf x}+{\bf x'})$ (blue),  $p({\bf y}+{\bf y'})$ (green) and $p({\bf z}+{\bf z'})$ 
(orange) for systems with $N^{\{4,1\}}=80k$ (left) and $N^{\{4,1\}}=$160k (right).  }
\label{shifts}
\end{figure}

Distributions of this type contain interesting information about geometric properties
of the system in spatial directions. We should remember that simplices with coordinates ${\bf x}=1$ and ${\bf x'}=1$ lie
on a {\it minimal} boundary between elementary cells. Such a boundary separates regions where
gravitational fluctuations produce volume concentrations. Qualitatively, in the small  ${\bf x}$ and ${\bf x'}$ regions
the volume of the system is suppressed.
A special role is played by the distribution $p(L_x)$ of the quantity $L_x = {\bf x}+{\bf x'}$ 
(and similar quantities $L_y$ and $L_z$).
Although for the initial configuration the shape of the elementary cell
is symmetric with respect to the exchange of directions ${\bf x}$, ${\bf y}$ and ${\bf z}$, during the  
thermalization process the shape of the elementary cell may be deformed. In fact, we expect that the distributions $p(L_x)$, $p(L_y)$ and $p(L_z)$
should overlap, up to a finite shift in $L_i$. This is indeed the case. In the figure \ref{shifts} we show
the distributions $p({\bf x}+{\bf x'})$, $p({\bf y}+{\bf y'})$ and $p({\bf z}+{\bf z'})$ for systems with $N^{\{4,1\}}=80k$
and $N^{\{4,1\}}=$160k. In both cases a complete overlap is achieved by applying shifts of  order one.

\begin{figure}[H]
\begin{center}
{\includegraphics[width=10cm]{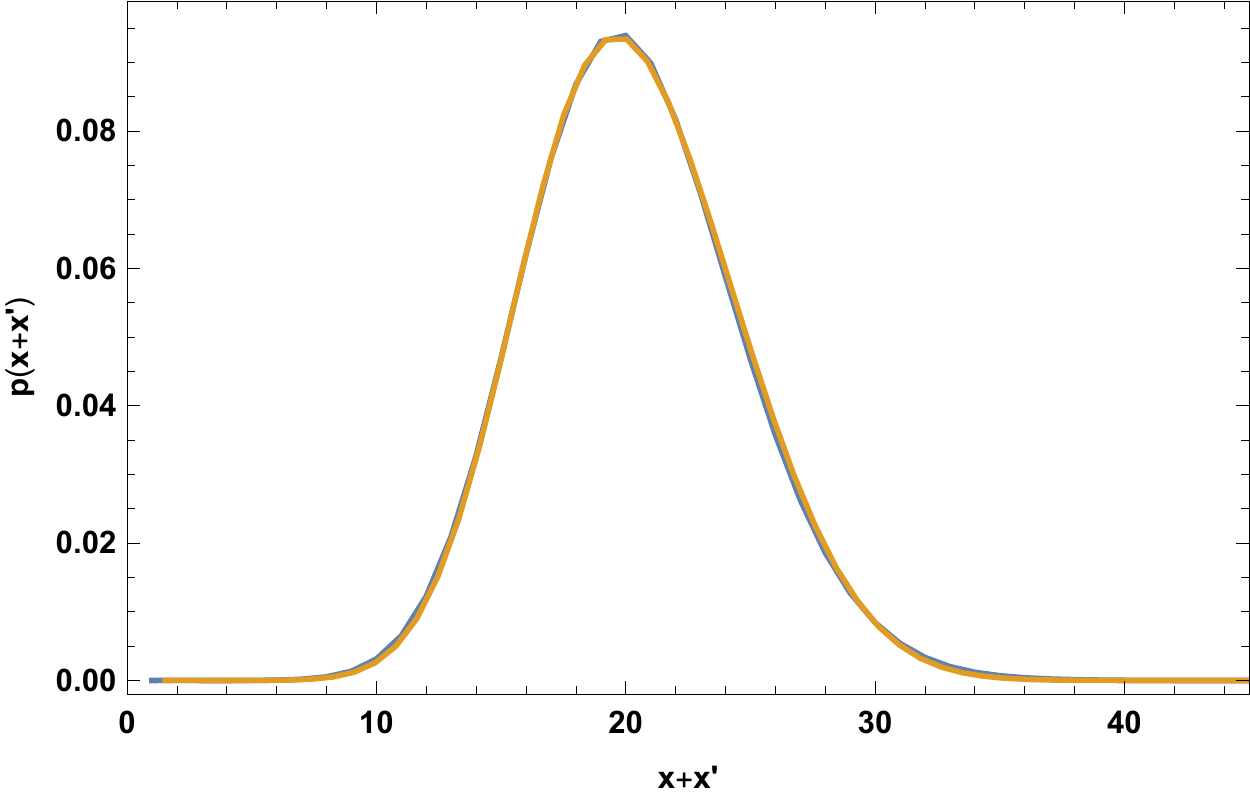}} 
\end{center}
\caption{The scaling of a distribution $p({\bf x}+{\bf x'})$ for two system sizes ($N^{\{4,1\}}=$80k and 160k,
colored  blue and orange, respectively).
For the larger system the distribution is rescaled by a factor $1/2^{1/4}$.}
\label{scaling}
\end{figure}

Comparing the two plots we realize that the shape of distributions seems to be the same, up to a scaling, 
depending on the total volume. In the figure \ref{scaling} we compare the two volumes ($N^{\{4,1\}}=$80k and 160k), applying a scaling factor $1/2^{1/4}$ to the 
distribution for a larger system. This agrees with the expected scaling, if we assume that the Hausdorff dimension
in de Sitter phase is $d_H=4$.

\begin{figure}[H]
\begin{center}
{\includegraphics[width=10cm]{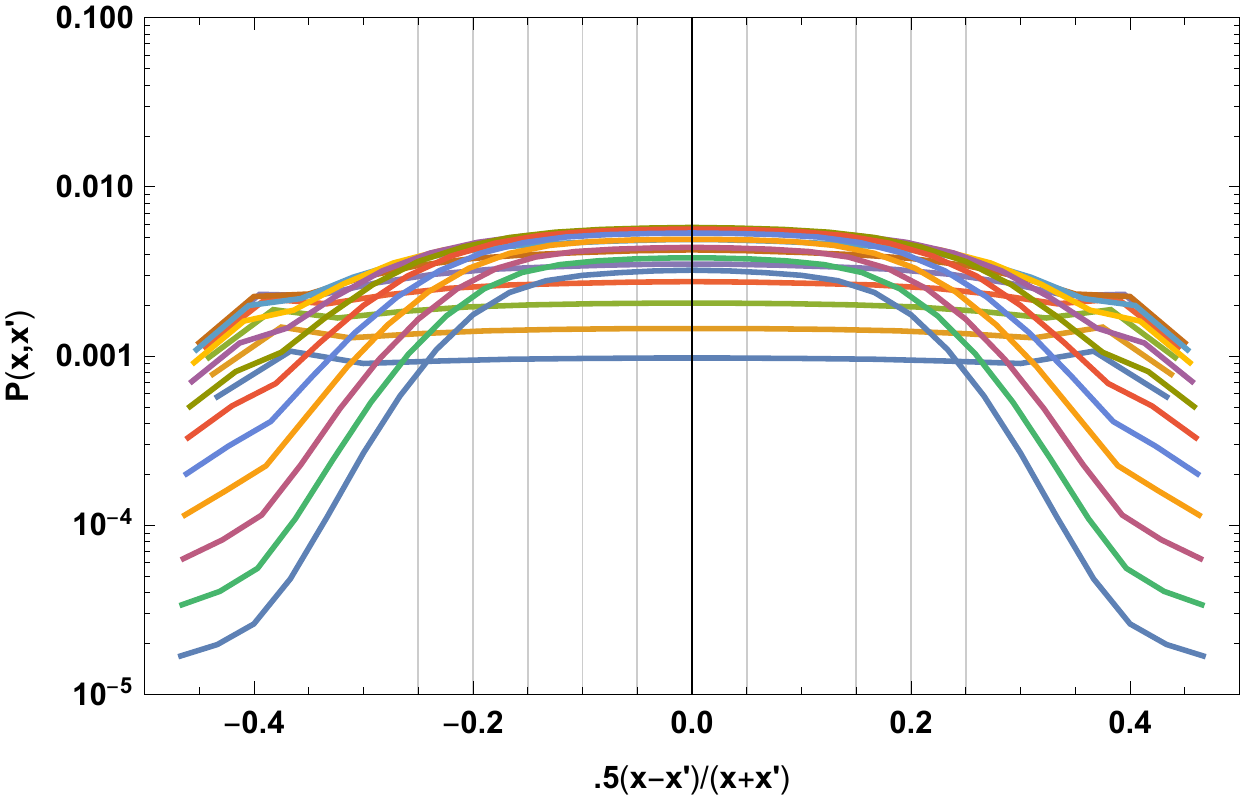}} 
\end{center}
\caption{Central part of the distribution $P({\bf x},{\bf x'})$ as a function
of the re-scaled variable $\oh({\bf x}-{\bf x'})/({\bf x}+{\bf x'})$. The curves correspond to the increasing values
of ${\bf x}+{\bf x'}$ in a sequence of colors: blue, orange, green etc.}
\label{middle}
\end{figure}

\begin{figure}[H]
\begin{center}
{\includegraphics[width=10cm]{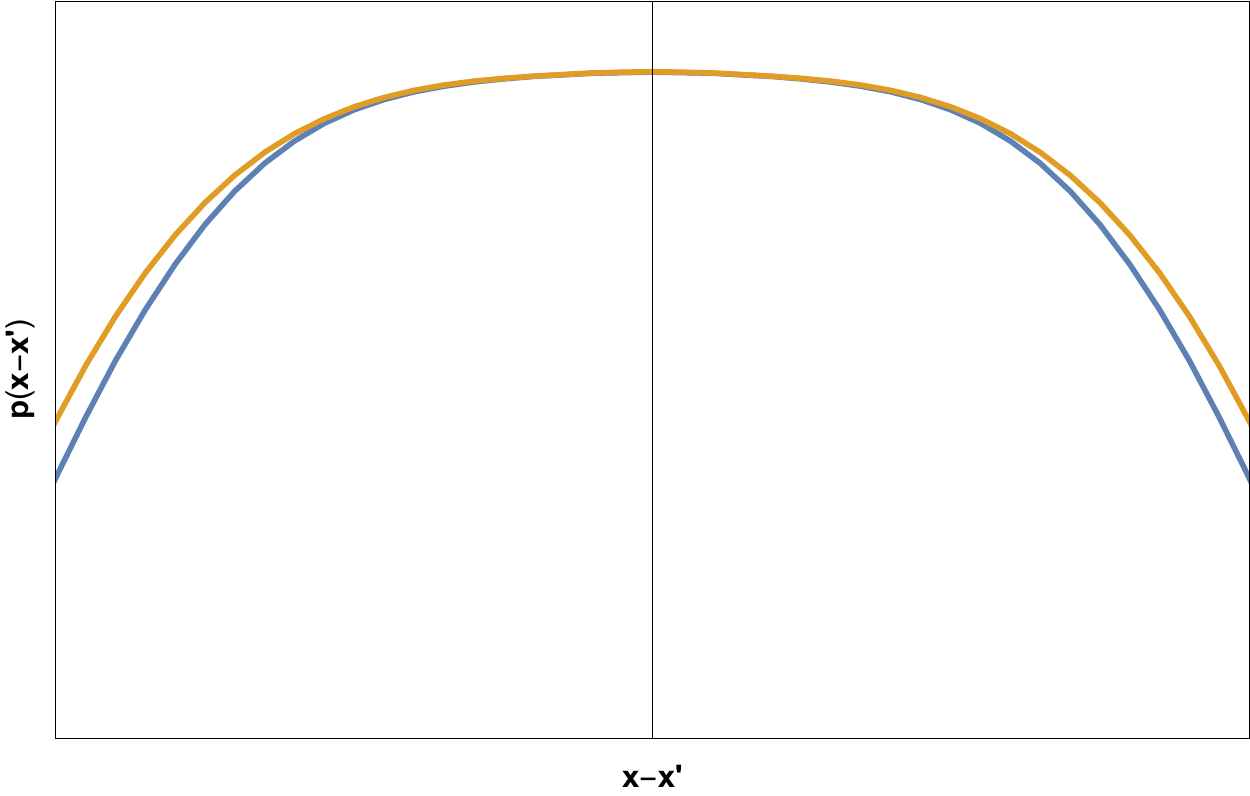}} 
\end{center}
\caption{The integrated distribution $p({\bf x}-{\bf x'})$ as a function
of ${\bf x}-{\bf x'}$ for volumes $N^{\{4,1\}}=$80k (blue) and 160k (orange).}
\label{middlescale}
\end{figure}

 Another important information about the shape of the volume distribution in spatial directions is hidden in the
distribution $p({\bf x}-{\bf x'})$ of the difference between ${\bf x}$ and ${\bf x'}$.
For a toroidal topology one expects this distribution to be approximately constant,
as was already  observed in  \cite{torus2} for a similar distribution in time $t$. While the periodicity in time 
was exact, the period in the spatial directions, say the ${\bf x}$ direction, will vary, 
depending on $L_x = {\bf x}+{\bf x'}$. As a consequence, we break the
spatial translational symmetry and force the system to have the smallest volume at the small ${\bf x}$ limit.
In the central range of the plot, where ${\bf x}\approx {\bf x'}$ we expect the
distributions to be approximately flat,  constant in the infinite volume limit.
Below, in the figure \ref{middle}, we show the amplified dependence of the volume distribution $P({\bf x},{\bf x'})$ 
as a function of the re-scaled variable $\oh({\bf x}-{\bf x'})/({\bf x}+{\bf x'})$ for a range of values of 
$15\leq {\bf x}+{\bf x'}\leq 30$ for a system with $N^{\{4,1\}}=$160k.
We see that the distributions  are indeed approximately flat in the central range. 
The large volume limit  can be extrapolated by comparing  distributions of $p({\bf x}- {\bf x'})$ for systems
with volumes 80k and 160k. We see that for a larger system the distribution gets wider (Figure \ref{middlescale}).

\section{The fractal structure of equal ${\bf x}$ layers}\label{sec:4}

As explained above, the constant ${\bf x}$ layer is by 
construction connected for ${\bf x} = 1$. We will study the
geometric structure of layers with higher values of ${\bf x}$. To do this
we define the concept of a {\it slice}. The idea is very similar to that used
in 2d Euclidean Dynamical Triangulations. The construction is based on 
following the front of a diffusion wave on the dual lattice. We start at a randomly chosen simplex with spatial coordinate ${\bf x}$ and constrain the diffusion process to take place in the subset of simplices with spatial coordinate ${\bf \tilde x} \geq {\bf x}$, i.e., in the bounded part of  
the given triangulation lying in between the set of simplices with  coordinate 
${\bf x}-1$ and the boundary of the elementary cell.
  In each diffusion step we mark simplices we meet on the way.
The marked simplices form a new limiting layer for the next step.

If the geometry of the manifold were trivial, 
in such a process we 
could reach all simplices in the bounded region of the configuration described above. 
The fractal nature of geometry means that, in general, only a part of the
bounded region is reached. Completing the process we find a set of 
simplices with the same coordinate ${\bf x}$ as the initial simplex.
The set of these simplices forms a {\it slice}. All simplices belonging to
the slice can be joined by at least one path lying in the bounded region.
Repeating the same process starting at a different simplex with a 
coordinate ${\bf x}$, we either end with the same slice or  produce
a different slice, disconnected from the previous one. The set of simplices
reached from the set belonging to a common slice has a tree-like structure,
where the slice plays a role of the root.

The procedure is performed repeatedly, assigning in this way
a slice index to all simplices of the configuration:
\begin{itemize}
\item The first slice consists of all simplices with ${\bf x}=1$.
\item In the layer with ${\bf x}=2$ we find all simplices that can be
connected by a path that does not cross the elementary cell boundary  and never goes below ${\bf x}=2$.
\item We find all separate slices for ${\bf x}=2$.
\item We repeat the same procedure for ${\bf x}=3$. In this case, we restrict the class of paths not to go below ${\bf x}=3$.
\item We continue until all slices are found.
\end{itemize} 
{For a (two-dimensional) visualization of the procedure please refer to Figure \ref{fig:visualize}.}
\newline

By construction, each slice (except the first one) has exactly one parent 
slice, and it may have a larger number of children (or no children at all). Following
the sequence of connected slices with an increasing value of ${\bf x}$, we 
may either reach the boundary or end in a blind alley. In the first 
case we say that the sequence belongs to the {\it trunk}, in the second case
it belongs to a {\it branch}. The structure looks like a tree, with the trunk
connecting the lower and upper boundaries. The trunk may split into
several outgrowths. The branches emerge from the trunk slices and, by definition, never 
reach the boundary of the elementary cell. 
We would like to interpret the branches as the effect of quantum 
fluctuations and the trunk as a semi-classical background.

\begin{figure}[H]
\begin{center}
{\includegraphics[width=7cm]{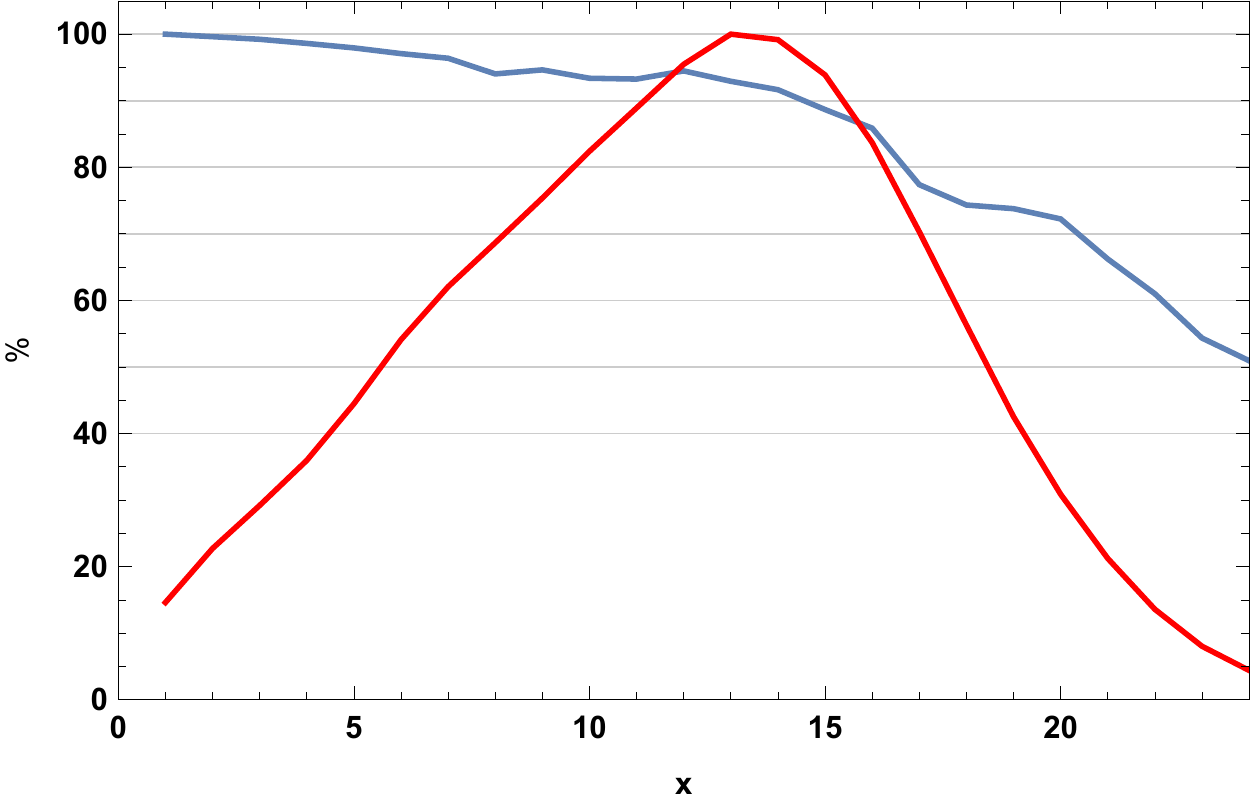}} 
{\includegraphics[width=7cm]{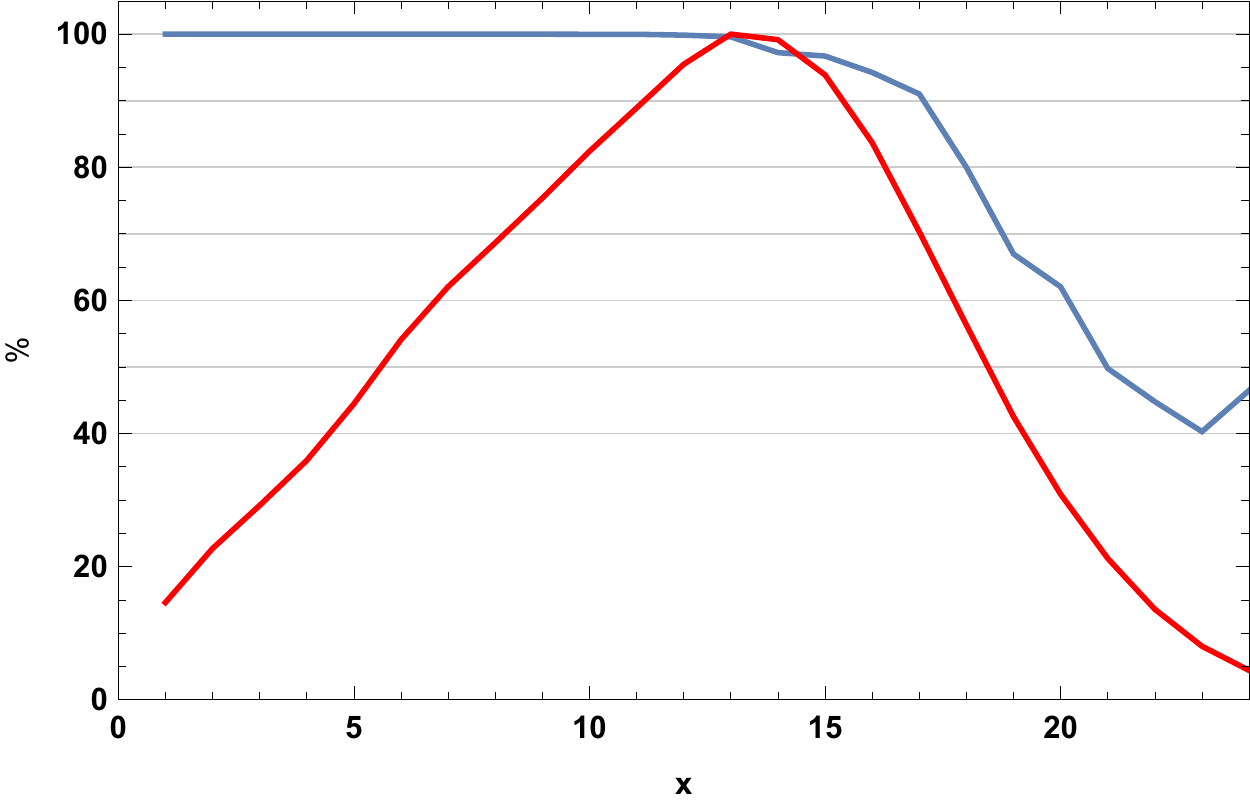}} 
\end{center}
\caption{The fraction of the trunk volume in the total volume (left) and of the largest slice volume in the
trunk volume (right). The (normalized) total volume distribution is drawn in red.}
\label{tree}
\end{figure}

Below we illustrate the properties of the trunk-branch construction for one 
particular well-thermalized configuration in the de Sitter phase. The 
configuration was obtained
for a system with $K_0=2.2$, $\Delta=0.6$, $T=4$ and the total number of 
simplices $N_4=370~724$. At each value of the coordinate ${\bf x}$ we split 
the volume into the part belonging to the trunk and the part belonging to branches. In the figure \ref{tree} we show the fraction (percent) of volume
in the trunk, relative to the total volume at a coordinate ${\bf x}$, and the fraction of volume of the largest slice, relative to the
trunk volume at the same coordinate ${\bf x}$. In the same plot (in red)
we show the total volume distribution as a function of ${\bf x}$. The 
distribution is normalized to reach $100\%$ at its maximum. The distribution in the plot (as compared to that in the figure  \ref{xxprim}) can be viewed as a
projection of volume on the ${\bf x}$ axis. In the plot all distributions are averaged over directions ${\bf x}$,
${\bf x'}$, ${\bf y}$, ${\bf y'}$, ${\bf z}$ and ${\bf z'}$.

As can be seen, for a configuration in the de Sitter
phase, the trunk contains the biggest part of volume
and the largest slice captures almost all trunk volume
up to the value of ${\bf x}$ where the volume is maximal (recall that by construction
the volume as a function of ${\bf x}$ is small near the boundaries).

\begin{figure}[H]
\begin{center}
{\includegraphics[width=7cm]{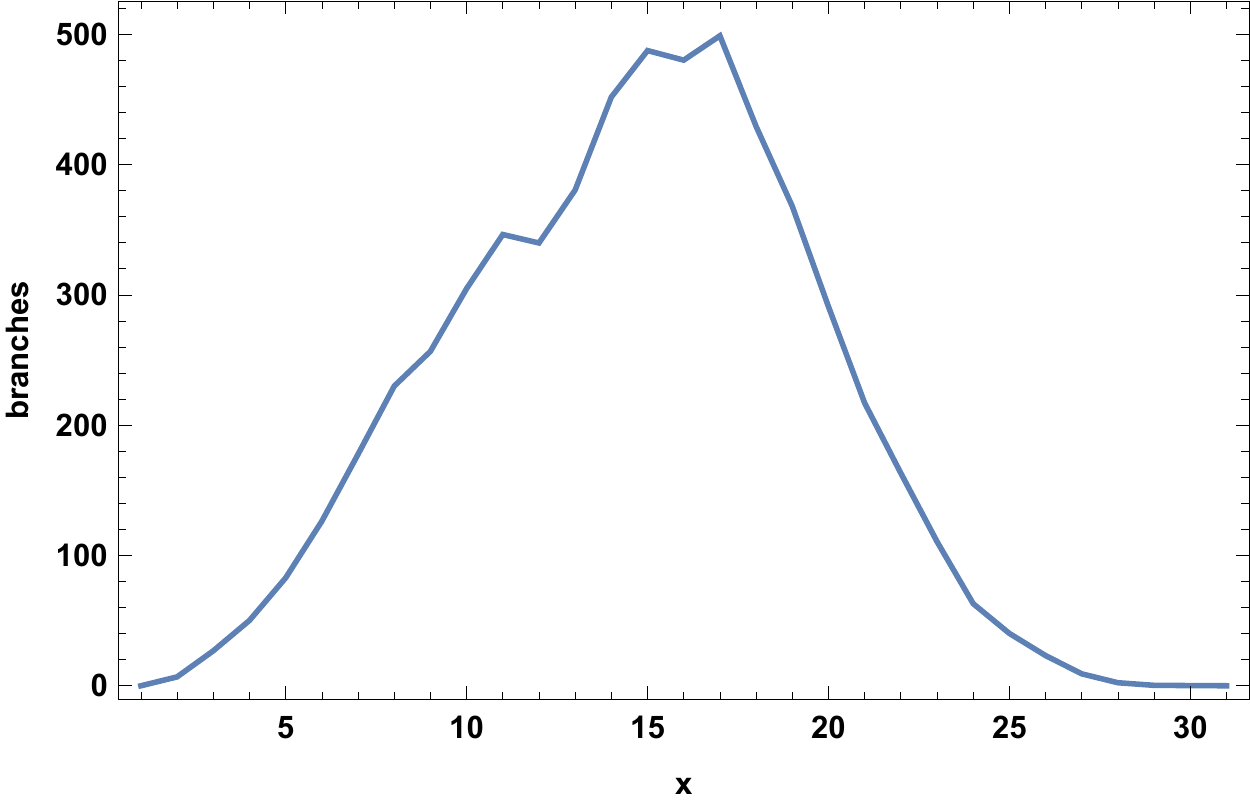}} 
{\includegraphics[width=7cm]{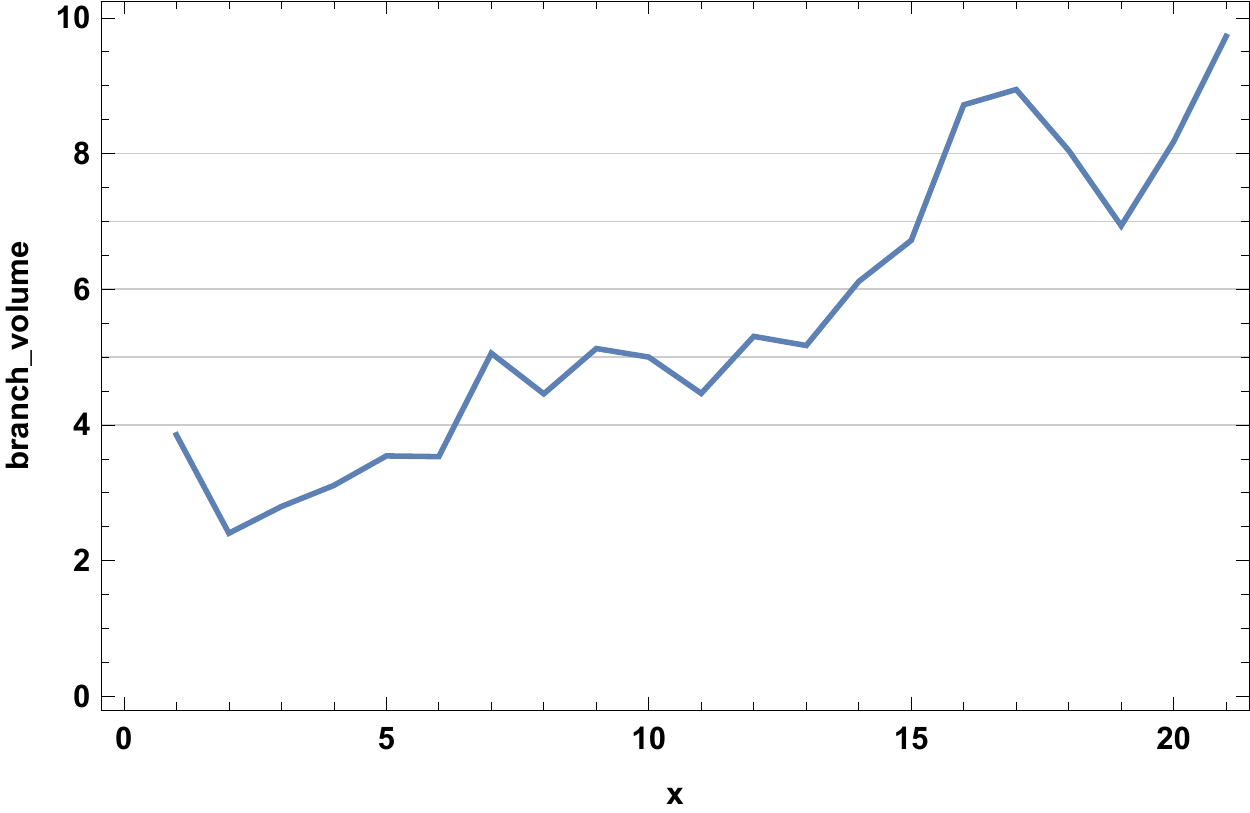}} 
\end{center}
\caption{The number of branches at a distance ${\bf x}$ (left) and the average branch volume (right). } 
\label{branches}
\end{figure}

In the next plot, figure \ref{branches}, we show the average number of branches 
at a distance ${\bf x}$ and the average volume per 
branch. As can be seen, the number of branch slices grows
together with  volume, but the average volume per branch slice
is relatively small. The maximal length of a branch, defined as 
a number of steps between the consecutive slices 
along the branch necessary to reach the trunk slice
from a branch slice, is six, which means that branches
are rather short.

\begin{figure}[H]
\begin{center}
{\includegraphics[width=14cm]{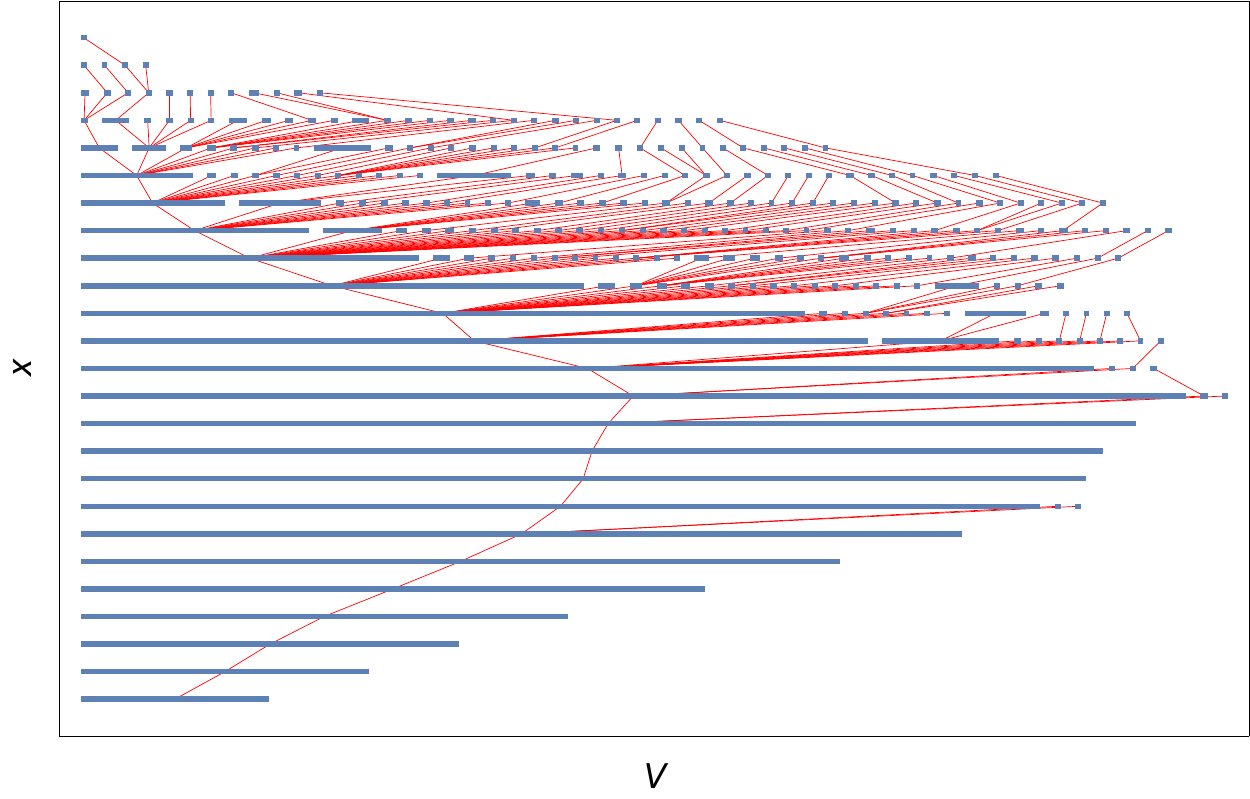}}
\end{center}
\caption{The tree structure of the branching for the increasing value of a coordinate ${\bf x}$. For clarity
 each slice is represented as a line with the length proportional to its volume. The red lines connect the centers of the
 consecutive slices.} 
\label{tree1}
\end{figure}

The tree structure representing the branching of slices with the increasing value of a coordinate ${\bf x}$ is
presented in the figure \ref{tree1}. For clarity the plot contains only the slices belonging to the trunk.
The volume of slices is shown as the horizontal length of the line.

\section{Loops with a non-trivial winding number}\label{sec:5}

In the former section we analyzed the fractal structure of an elementary cell, parametrized by what we denoted as
pseudo-Cartesian coordinates. Using the same elementary cell and its periodic extensions to neighboring cells,
we will now introduce a set of new topological observables, which measure the geodesic distance between a simplex
and its copy in the neighboring cells in different spatial directions. 
Like before, the measurement is performed by following paths defined by the front of
a diffusion wave, starting from a given simplex, but
in this case we consider the torus as an infinite system where simplices
reached after $r$ diffusion steps are labeled by the (unique) index in the elementary cell and also by the number of times a path crossed the 
boundaries. If one finds a simplex with the same cell index and nontrivial number of crossings, say, 
$\{m_1,~m_2,~m_3,~m_4\}$, it means that there exists at least one topologically closed geodesic line characterized by these  winding numbers, namely the shortest such path\footnote{The ``shortest'' is meant in a graph-theoretical sense: we follow a 
link path (on the dual lattice) and the length is the number of links. This is not necessarily the shortest path 
if we consider our triangulation as a piecewise linear manifold. However, we do not expect this difference 
to be important when we discuss generic fractal properties, like the Hausdorff dimension of a generic triangulation
in the limit of infinitely large triangulations. By an abuse of language we still call the shortest link path a geodesic.}.
In general there are many paths with the same length, but for a particular simplex and particular values
$\{m_1,~m_2,~m_3,~m_4\}$ there is always the shortest distance, characterizing these two copies of a simplex. 
In a $4D$ system the number of elementary cells 
at a large distance $R$ is expected to grow as $R^3$.
As a consequence, measuring the complete distribution of paths up to some length $R$ is computationally
 very time-consuming, therefore we decided to perform  measurements for a single well-thermalized
 configuration with $N^{\{4,1\}}=$160k (the total number of simplices in this configuration is $N_4 = 370~724$) 
 and the time periodicity $T=4$ (in fact this is the same configuration we used in the preceding section).
 For this configuration we measured loops in all spatial and time directions, restricting the length to be below 55 
 (in this way we could measure the distribution of loop geodesics to the nearest copies in all directions and additionally
 some geodesics with more complicated winding characteristics). Although the analysis
 is based only on one configuration, we expect that distributions will self-average to produce a reasonable
 estimate for a whole ensemble. 
  
\begin{figure}[H]
\begin{center}
{\includegraphics[width=10cm]{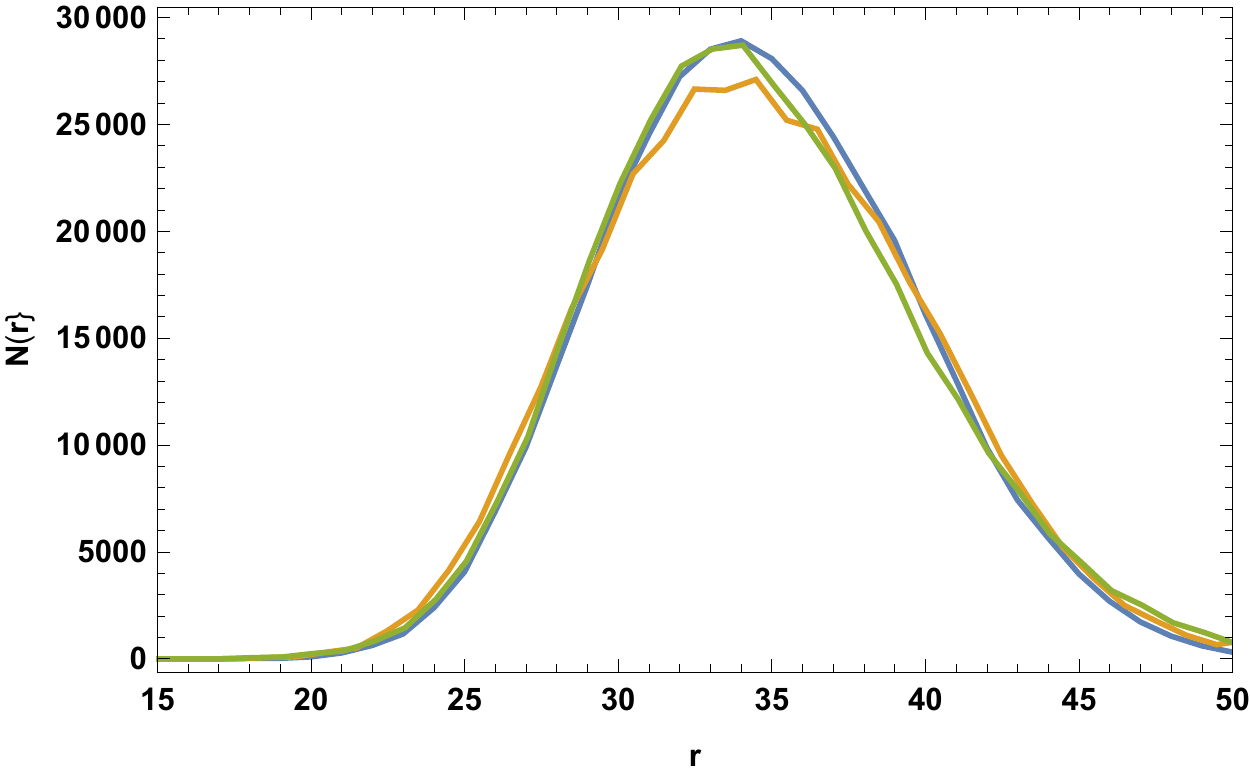}} 
\end{center}
\caption{Distributions of loop distances for loops with winding numbers $\{1,~0,~0,~0\}$ (blue), $\{0,~1,~0,~0\}$ (orange) and $\{0,~0,~1,~0\}$ (green) shifted in $r$ by a shift of order 1.}
\label{wind}
\end{figure}

We expect the distributions of lengths of all loops connecting the neighboring cells  in $x$, $y$ and $z$ directions to be approximately the same,
up to a possible small shift in the length $r$. This is indeed what is observed. The figure
\ref{wind} shows the length distributions of the $\{1,~0,~0,~0\}$, $\{0,~1,~0,~0\}$ and $\{0,~0,~1,~0\}$ 
loops superimposed using a shift of order smaller than 1. 
The match is not perfect, but we should remember that we are using just a single configuration. The need of making small finite
shifts reflects the fact that during the thermalization process the shape of the elementary cell is deformed from the 
regular symmetric shape it had in the initial hypercube. 

\begin{figure}[H]
\begin{center}
{\includegraphics[width=10cm]{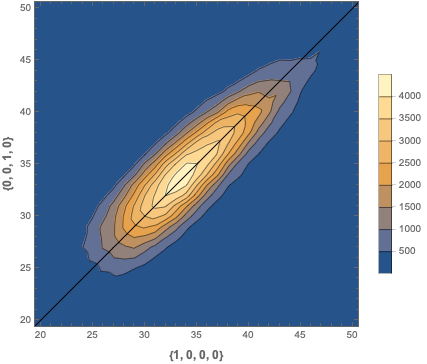}} 
\end{center}
\caption{Correlation between loop distances for loops with winding numbers $\{1,~0,~0,~0\}$  and $\{0,~0,~1,~0\}$.}
\label{loop13}
\end{figure}

Comparing the loop distance in two different directions (Figure \ref{wind}), we
find that not only is the shape of the distributions similar. 
In fact loop distances for {\it all simplices} are highly 
correlated, which can be seen in the figure \ref{loop13}, which presents the correlation between the loop 
distance in the $\{1,~0,~0,~0\}$ and $\{0,~0,~1,~0\}$ directions.

For all simplices in the configuration we assigned a set of numbers measuring distances to their copies in the neighboring cells.
These numbers contain interesting information about the internal geometry of the manifold. 
\begin{itemize}
\item In each direction we 
find a finite number of loops with the minimal length $r_{min}$.
The minimal loops follow the deepest {\it valleys} being surrounded by {\it mountains} which we can view as   
large volume fluctuations. In some sense, the minimal loops are dual to 
the minimal boundaries discussed in previous sections.
\item The simplices located on the {\it mountains} (i.e., inside the large volume fluctuations) 
are characterized by longer loop distances to the neighboring cells.
Given a simplex, we can compare its loop
distance in each direction with 
the loop distances of
its five neighbors.
 For all simplices with the same loop distance $r$, we find
that the neighboring simplices are characterized by loop distances in the range  $r$, $r\pm 1$ and $r\pm 2$. 
\item Only for the few shortest loops can we
follow a simple path connecting the simplex to its copy that would go through 
a sequence of neighboring simplices
characterized by the {\it same} value of the loop distance $r$. 
\item For longer loops all paths to a copy pass through fragments
of loops with smaller loop distances. 
\item We deduce that the difference between the loop distance $r$ for a particular simplex and the minimal distance 
$r_{min}$ indicates how far a given simplex is from the minimal loop.
\end{itemize}

\begin{figure}[H]
\begin{center}
{\includegraphics[width=10cm]{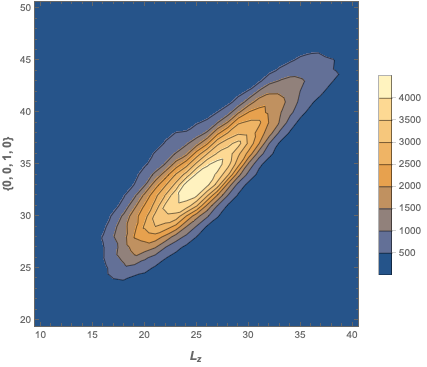}} 
\end{center}
\caption{Correlation of a distribution of $L_z$ (horizontal axis) with a loop distance in $\{0,~0,~1,~0\}$ direction.}
\label{zvs001}
\end{figure}

Finally, we will compare the distribution of loop distances for simplices in the fundamental cell in, say, the $\{0,~0,~1,~0\}$ direction to
the distribution of the quantity $L_z={\bf z}+{\bf z'}$ discussed earlier.
We recall that this quantity
is the sum of the distances from a simplex to the two cell boundaries {\it orthogonal} to the ${\bf z}$ direction.
We find that the two distributions are highly correlated. An example of this correlation 
can be seen on the contour plot shown in the figure \ref{zvs001}.

\section{Loop distances in other directions}

\begin{figure}[H]
\begin{center}
{\includegraphics[width=12cm]{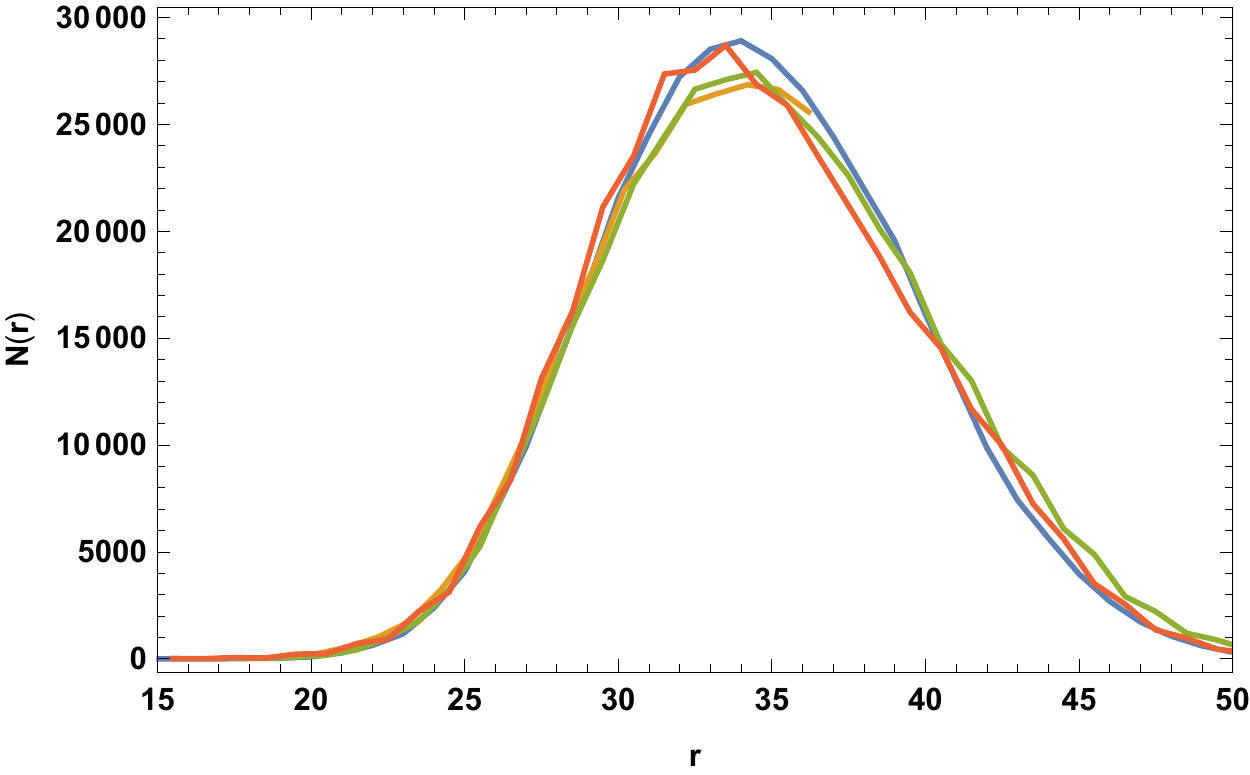}} 
\end{center}
\caption{Distributions of (shifted) loop distances  in directions $\{1,~0,~0,~0\}$ (blue), $\{1,~-1,~0,~0\}$ (red), $\{1,~-1,~1,~0\}$
(green) and $\{1,~1,~1,~0\}$ (orange) showing  the universal blob structure.}
\label{collected}
\end{figure}

The universality of the loop distance distribution can be extended to include loops with higher non-trivial winding numbers.
The interpretation is that the blob-like structure (valleys and mountains) reflects the fractal structure of the fluctuations
of geometry in the model and therefore can be expected to be independent of the direction in which we 
measure the loop distance distribution. 
For different directions we expect different minimal loop lengths, and as a consequence of comparing 
distributions with different argument range we may have to adjust this range to obtain a proper overlap 
when we compare the distributions.
 In Figure \ref{collected} we show the comparison of distributions for
loop distances in directions $\{1,~0,~0,~0\}$ (blue), $\{1,~-1,~0,~0\}$ (red), $\{1,~-1,~1,~0\}$
(green) and $\{1,~1,~1,~0\}$ (orange).  The last one is seen
only as a fragment of a distribution, 
as it requires a shift of 15 units, and is not contained fully in the region $R<55$ we analyzed. The universality of the distributions, independent 
of directions, strongly suggests that the distribution of volume fluctuations is spherically symmetric. 
At the same time, the difference in the minimal loop distance in different directions most likely 
reflects that the shape of the elementary cell of our thermalized configuration is non-trivial.

\begin{figure}[H]
\begin{center}
{\includegraphics[width=12cm]{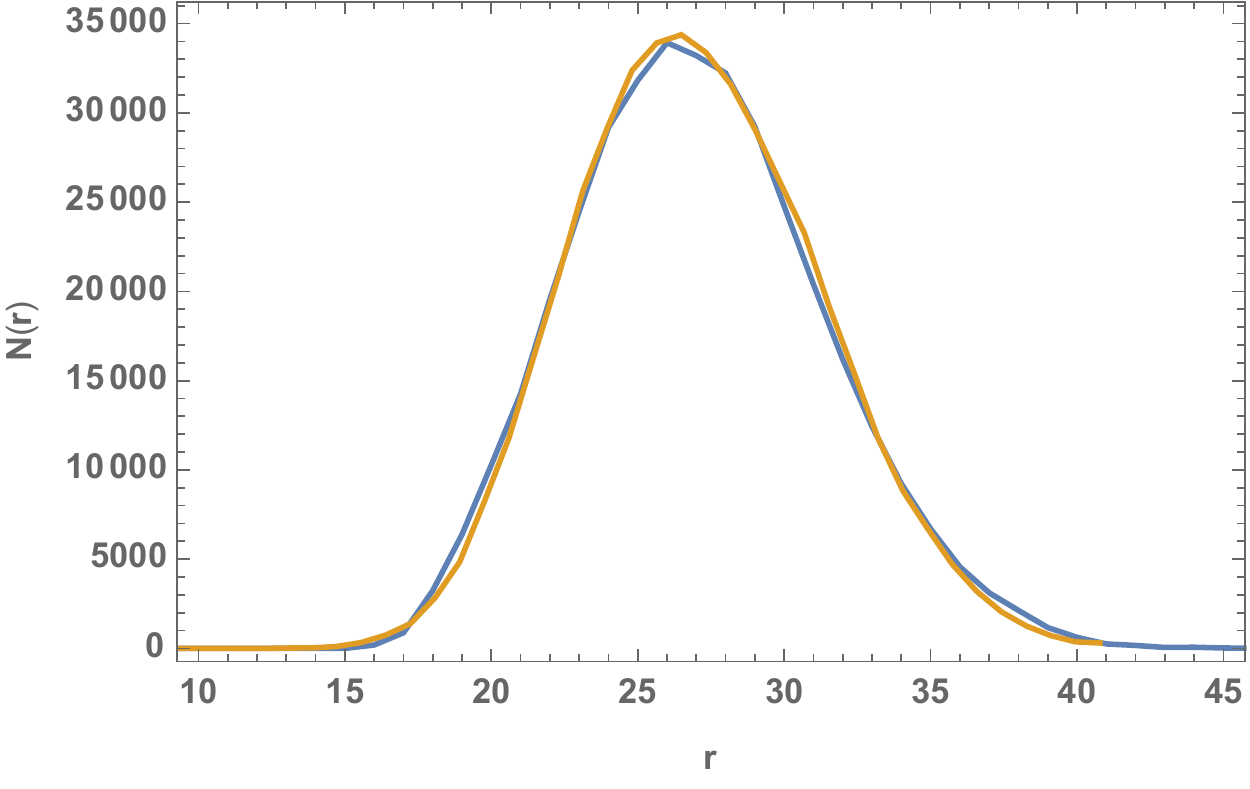}} 
\end{center}
\caption{The loop distance distribution in the time direction, compared to that in the ${\bf x}$ direction.}
\label{loopt}
\end{figure}

The thermalized configuration discussed here has the four-dimensional topology of $T^4$. 
In the time direction there is
a foliation, which is absent in spatial directions. Although the periodicity in the time direction is trivial,
this does not necessarily mean that the distribution of the loop distances in the time direction $\{0,~0,~0,~1\}$ will
be a Dirac delta function. In fact, due to the large spatial volume fluctuations, it has a shape very similar to that in spatial directions. Below,
in the figure \ref{loopt}, we show the distribution of the loop distances in the time direction ($\{0,~0,~0,~1\}$) (blue)
as compared to a shifted and rescaled distribution in the $\{1,~0,~0,~0\}$ direction (orange). The
scaling factor used was 1.19. 

This result indicates that although we have a foliation in time, the information encoded in the loop distance
in spatial and time directions is very similar. The average length of the loop distance in the time direction is
expected to depend on the time period $T$.  

\section{Conclusion}
 
 CDT is a non-perturbative quantum gravity model. It is a model with a build-in cut-off,
 the length  $a$ of the links in the triangulations. In order to prove that it is an interesting model of 
 quantum gravity one has to be able to remove the cut-off in a controlled way and show that the resulting 
 theory satisfies the criteria  suitable for a theory of quantum gravity\footnote{A discussion of how
 the CDT cut-off can in principle be removed while physics is kept constant can be found 
 in \cite{rge}.}. Two such criteria often 
 mentioned are background independence and diffeomorphism invariance. CDT partially satisfies these 
 criteria already at the regularized level when the cut-off $a$ is non-zero. It is background independent 
 in the sense that it is defined as a sum over a certain class of piecewise linear geometries, and each geometry 
 is given the same weight, except for the weight associated with the Hilbert-Einstein action. However, more 
 remarkable is the fact that the path integral is carried out explicitly as a sum over geometries. The analogy
 in a gauge theory would be performing the path integral directly over the set of gauge equivalent 
 field configurations. This possibility comes about because we regularize our path integral by using a certain 
 class of piecewise linear geometries, and, as emphasized by Regge \cite{regge}, one does not need coordinates
 in order to describe the geometry of piecewise linear geometries. 
 In this sense, not having coordinates is very 
 good and in some sense consistent with a central point in general relativity, where physics is invariant under coordinate transformations.
 However, coordinates can be very useful, in particular the {\it right} coordinates, for the description of a certain 
 physical phenomenon. In \cite{four-d} we showed that using a proper-time coordinate the ``quantum universe'' 
 observed in the computer simulations where space had the topology of $S^3$ could be described by 
 a semi-classical effective action. However, we also encountered the problem that it was difficult 
 to extend the effective action to include space coordinates. Our representation was simply too 
 ``diffeomorphism invariant''.  If the topology of space is that of $T^3$, the situation is somewhat better. 
 As we have described above, one has the possibility to use the periodic nature of $T^3$ to construct a kind of 
 pseudo-Cartesian spatial coordinate system.  Its status is however somewhat different from the way a 
 coordinate system usually appears in a quantum gravity calculation where the geometry is described 
 in terms of $g_{\mu \nu}(x,y,z,t)$, with $x,y,z,t$ referring to a given coordinate system, the same for any
 $g_{\mu \nu}(x,y,z,t)$ which appear in the path integral. In our case, we are in principle adjusting our 
 coordinate system to the geometry. The usefulness of such a construction will depend on the nature
 of the quantum system. If the quantum system is dominated by one kind of configurations, on top of which 
 there are ``not-too-wild'' quantum fluctuations, this construction can be very useful, in particular if the dominating 
 configuration is not a simple classical or semiclassical configuration. This is seemingly the situation we 
 meet in the de Sitter phase of our quantum gravity model. Certain aspects of the dominating configurations 
 have a semiclassical interpretation in the sense that their features can be described by a suitable time coordinate
 and an {\it effective} action depending on the time coordinate. But this action is far from any $\hbar$ expansion
 used in the path integral and thus far from an expansion related to solutions of 
 the classical action. For such an {\it effective} action it is {\it a priori} not clear how to choose a good coordinate
 system. For this reason we have chosen the ``dynamical'' approach described above, where the fundamental
 cell of  $T^3$  is chosen to be as regular as possible for a given configuration. In terms of the ``pseudo-Cartesian''  
 coordinates associated with the chosen cell, we observe 
 an interesting spatial structure: a ``trunk'' of connected four-simplices for each spatial direction, decorated by 
 quantum outgrowths. 
 This suggests a ''semiclassical'' interpretation even of a spatial section of the four-dimensional
 universe, which is not so different from the situation in two-dimensional Euclidean quantum gravity, where
 a typical configuration, when sliced appropriately (choosing the appropriate coordinates for the configuration 
 in question), can be viewed as a main universe dressed with quantum outgrowths (baby universes) \cite{kawai}. 
 In the case of two-dimensional Euclidean quantum gravity, this slicing and its associated fractal structure 
 determines most aspects of two-dimensional Euclidean quantum gravity. Our preliminary analysis (only based 
 on one large configuration) suggests that the fractal structure of the three-dimensional configurations is 
 less fractal (i.e., more ``semiclassical'') than in the two-dimensional case, in the sense that there are fewer 
 outgrowths, and they carry less volume than in two-dimensional quantum gravity. Clearly,
 the precise determination of the fractal structure of the three-dimensional quantum space, such as the determination
 of critical exponents associated with this structure, requires considerably longer Monte Carlo simulations.
 However, the less fractal structure also increases the hope that in this way one can obtain a more 
 complete semiclassical effective action expressed in terms of the proper time and these new spatial coordinates. 
 
 The use of toroidal topology also allowed us to use loops with non-trivial winding numbers to extract information
 about the fractal structure of our spatial geometries. This has been done successfully in two-dimensional 
 Euclidean quantum gravity \cite{ab}, and also here in the four-dimensional case it provides us 
 with a lot of information. In some sense, the shortest non-contractible loops can be considered as dual 
 to the minimal cell boundaries and, like in 
 the two-dimensional case, the shortest loops lie in narrow ``valleys'', where the surrounding ``hills''  look like
 (topologically) spherical outgrowths. Again, further computer simulations should allow one to determine the fractal structure
 related to ``valleys'' and ``hills'', much in the spirit of  two-dimensional Euclidean quantum gravity.

 \section*{Acknowledgments} 
 
JA wishes to acknowledge support from the Danish Research Council grant ``Quantum Geometry''. JGS wishes to acknowledge support of the grant UMO-2016/23/D/ST2/00289 from the National Science Centre, Poland. ZD acknowledges a support of the
PhD project ``Kartezjusz''.
AG acknowledges support by the National Science Centre, Poland under grant 2015/17/D/ST2/03479. 



\begin{thebibliography}{99}

\bibitem{torus1}
J. Ambjorn, J. Gizbert-Studnicki, A. G\"{o}rlich, J. Jurkiewicz, D. N\'{e}meth,
{\it The phase structure of Causal Dynamical Triangulations with toroidal spatial topology},
JHEP 1806 (2018) 111, arXiv:1802.10434[hp-th];

\bibitem{report}
J. Ambjorn, A. G\"{o}rlich, J. Jurkiewicz, R.Loll,
{\it Nonperturbative Quantum Gravity},
Phys.Rept. 519 (2012) 127-210, arXiv:1203.3591[hep-th];


\bibitem{regge}
T.~Regge,
{\it General relativity without coordinates,}
Nuovo Cim.\  {\bf 19} (1961) 558.

\bibitem{two-d}

J.~Ambjorn and R.~Loll,
  {\it Nonperturbative Lorentzian quantum gravity, causality and topology change,}
  Nucl.\ Phys.\ B {\bf 536} (1998) 407, arXiv:
  [hep-th/9805108].\\
  J.~Ambjorn, R.~Loll, W.~Westra and S.~Zohren,
 {\it Putting a cap on causality violations in CDT,}
  JHEP {\bf 0712} (2007) 017, arXiv:
  [arXiv:0709.2784 [gr-qc]].\\
  J.~Ambjorn, R.~Loll, Y.~Watabiki, W.~Westra and S.~Zohren,
  {\it A String Field Theory based on Causal Dynamical Triangulations,}
  JHEP {\bf 0805} (2008) 032, arXiv:
  [arXiv:0802.0719 [hep-th]].
  
 \bibitem{three-d}
 
 J.~Ambjorn, J.~Jurkiewicz, R.~Loll and G.~Vernizzi,
  {\it Lorentzian 3-D gravity with wormholes via matrix models,}
  JHEP {\bf 0109} (2001) 022, arXiv:
  [hep-th/0106082];
  {\it 3-D Lorentzian quantum gravity from the asymmetric ABAB matrix model,},
  Acta Phys.\ Polon.\ B {\bf 34} (2003) 4667, arXiv:
  [hep-th/0311072].\\
  J.~Ambjorn, J.~Jurkiewicz and R.~Loll,
  {\it Renormalization of 3-d quantum gravity from matrix models,}
  Phys.\ Lett.\ B {\bf 581} (2004) 255, arXiv:
  [hep-th/0307263].
  
 \bibitem{four-d} 
 
 J.~Ambjorn, J.~Jurkiewicz and R.~Loll,
  {\it Emergence of a 4-D world from causal quantum gravity,}
  Phys.\ Rev.\ Lett.\  {\bf 93} (2004) 131301, arXiv:
  [hep-th/0404156];
  {\it Reconstructing the universe,}
  Phys.\ Rev.\ D {\bf 72} (2005) 064014, arXiv:
  [hep-th/0505154].\\
  J.~Ambjorn, A.~Gorlich, J.~Jurkiewicz and R.~Loll,
  {\it Planckian Birth of the Quantum de Sitter Universe,}
  Phys.\ Rev.\ Lett.\  {\bf 100} (2008) 091304, arXiv:
  [arXiv:0712.2485 [hep-th]];
  Phys.\ Rev.\ D {\bf 78} (2008) 063544, arXiv:
  [arXiv:0807.4481 [hep-th]].

\bibitem{torus2}
J. Ambjorn, Z. Drogosz, J. Gizbert-Studnicki, A. G\"{o}rlich, J. Jurkiewicz, D. N\'{e}meth,
{\it Impact of topology in causal dynamical triangulations quantum gravity},
Phys.Rev. D94 (2016) 044010, arXiv:1604.08786[hep-th];\\
J. Ambjorn, J. Gizbert-Studnicki, A. G\"{o}rlich, K. Grosvenor, J. Jurkiewicz, 
{\it Four-dimensional CDT with toroidal topology},
Nucl.Phys. B922 (2017) 226-246, arXiv:1705.07653[hep-th];

\bibitem{cube}
P.S. Mara. {\it Triangulations for the Cube},
Journal of Combinatorial Theory (A)20:170-177, 1976;

\bibitem{rge}
J.~Ambjorn, A.~Görlich, J.~Jurkiewicz, A.~Kreienbuehl and R.~Loll,
  {\it Renormalization Group Flow in CDT,}
  Class.\ Quant.\ Grav.\  {\bf 31} (2014) 165003, arXiv:1405.4585 [hep-th].

\bibitem{kawai}
  H.~Kawai, N.~Kawamoto, T.~Mogami and Y.~Watabiki,
  {\it Transfer matrix formalism for two-dimensional quantum gravity and fractal structures of space-time,}
  Phys.\ Lett.\ B {\bf 306} (1993) 19, arXiv:
  [hep-th/9302133].
  
  \bibitem{ab}
  
  J.~Ambjørn and T.~Budd,
  {\it The toroidal Hausdorff dimension of 2d Euclidean quantum gravity,}
  Phys.\ Lett.\ B {\bf 724} (2013) 328, arXiv:
  [arXiv:1305.3674 [hep-th]];
  {\it Semi-classical Dynamical Triangulations,}, 
  Phys.\ Lett.\ B {\bf 718} (2012) 200, arXiv: 1209.6031 [hep-th]].\\
  J.~Ambjorn, J.~Barkley and T.~G.~Budd,
  {\it Roaming moduli space using dynamical triangulations,}
  Nucl.\ Phys.\ B {\bf 858} (2012) 267, arXiv: 1110.4649 [hep-th]].\\
  J.~Ambjorn, J.~Barkley, T.~Budd and R.~Loll,
  {\it Baby Universes Revisited,}
  Phys.\ Lett.\ B {\bf 706} (2011) 86,
  arXiv: 1110.3998 [hep-th].

\end{thebibliography}
\end{document}